\documentclass[10pt,aps,twocolumn,prx,floatfix]{revtex4-2}
\usepackage{graphicx} 
\usepackage{xcolor}
\usepackage{amssymb,amsfonts,amsmath,amsthm}
\usepackage{bm}  
\usepackage{epstopdf}

\usepackage[hidelinks]{hyperref}
 \hypersetup{
     colorlinks,
     linkcolor={red!50!black},
     citecolor={blue!50!black},
     urlcolor={blue!80!black}
 }

\begin{document}

\title{DNA Unzipping Transition}

        \author{Somendra M. Bhattacharjee}
        \email{somendra.bhattacharjee\string@ashoka.edu.in}
        \affiliation{Department of Physics, Ashoka University, Sonepat,  Haryana-131029, India}

\begin{abstract}
  { This review focuses on the force-induced unzipping transition of
    double-stranded DNA. It begins with a brief history of DNA
    melting, which emerged alongside the growth of the field of
    molecular biology, juxtaposed with the advancements in physics
    during the same post-World War II period. The earlier theories of
    melting of DNA were based on the Ising model and its
    modifications, but gradually moved toward polymer-based models.
    The idea of force-induced unzipping was first introduced in 1999
    as a cooperative mechanism for breaking base pairs without the
    need for temperature changes.  The paper discusses several
    subsequent developments addressing different aspects of the
    unzipping transition.\\[8pt]
    {\it For EPJ B  Topical issue: ``100 glorious years of the Ising model"}.}
\end{abstract}


\maketitle


\newcommand{\milst}{%
  \begin{figure}[htbp]
    \centering
\includegraphics[width=1.0\linewidth,angle=0,trim=75 10 70 0,clip]{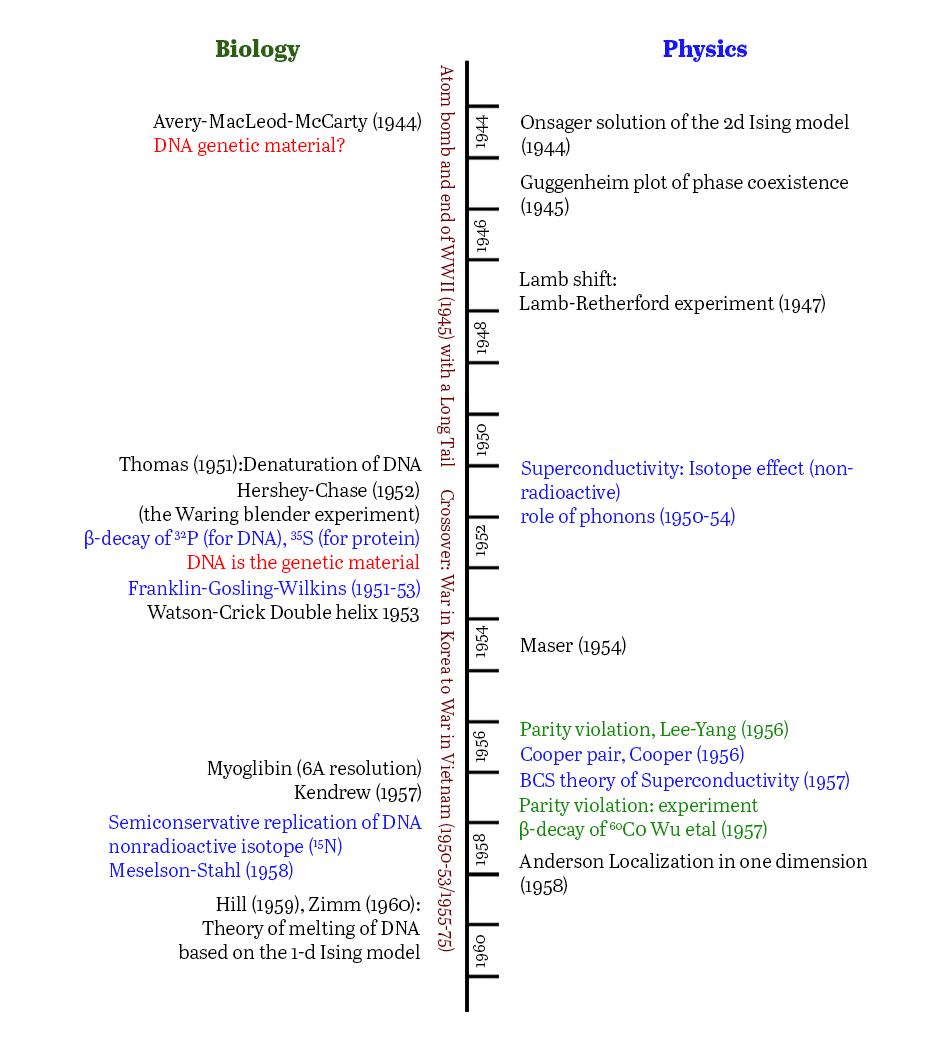}    
    
\caption{The 15-year period that saw significant breakthroughs in biology and physics.}
   \label{fig:1}
  \end{figure}
}%
\newcommand{\psbubble}{%
  \begin{figure}[hb]
    \centering
    \includegraphics[width=0.7\linewidth]{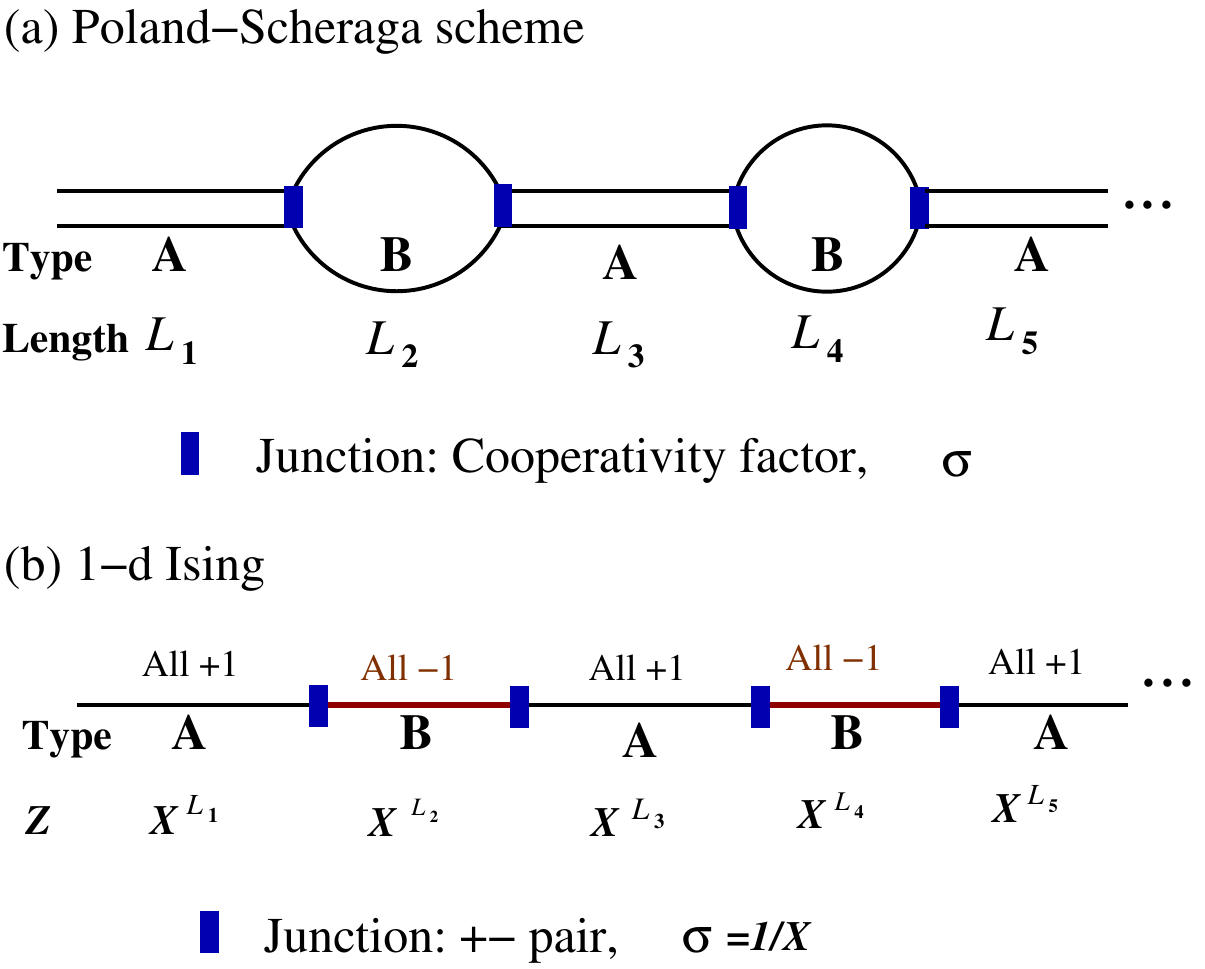}
    \includegraphics[width=0.7\linewidth]{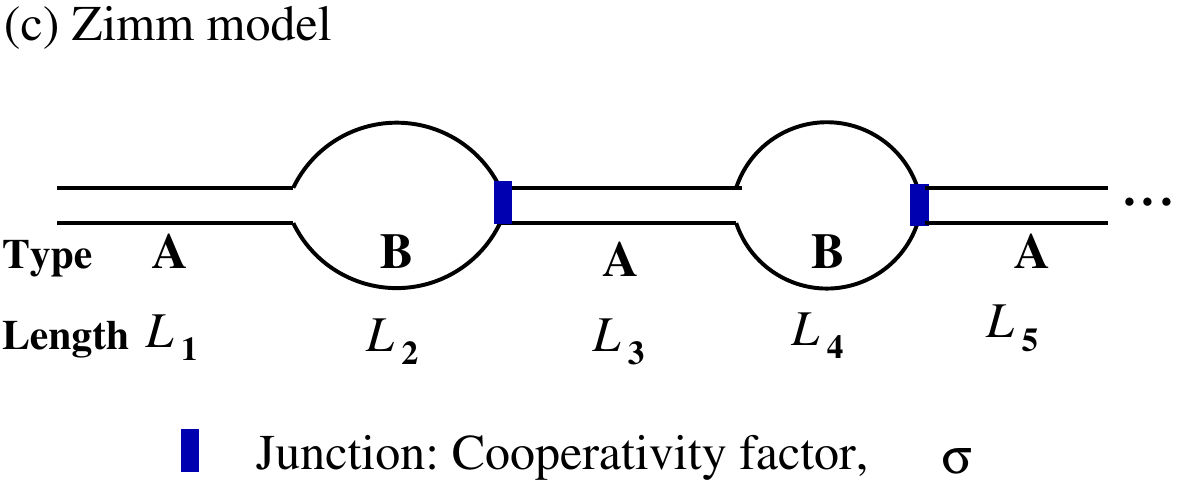}
    \caption{Typical configurations consisting of alternating sequnce
      of regions A and B.  The blue boxes indicate  junction points
      each of which contribute a weight $\sigma$ to the term in the partition
      function. (a) The Poland--Scheraga scheme for the melting
      transition. A$\equiv$ bound, B$\equiv$ unbound. (b) One
      dimensional Ising model in the Poland--Scheraga Scheme. A and B
      domains correspond to $+1$ and $-1$ spin domains, while the
      junction points are $\pm 1$ pairs. (c) The Zimm model where the
      special weight for the junction point is for unbound strands to join.} 
    \label{fig:psbub}
  \end{figure}
}%

\newcommand{\sshape}{%
  \begin{figure}[htbp]
    \centering
    \includegraphics[width=\linewidth,trim=10pt 60pt 0pt 55pt,clip]{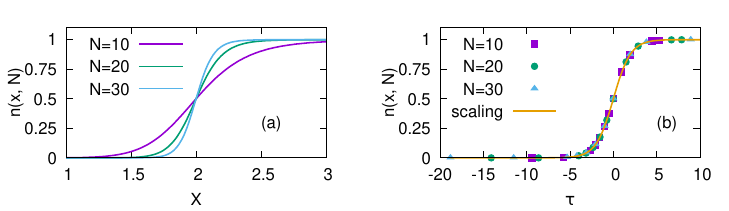}
   
    \caption{ The melting profile and finite-size scaling for $\mu=2$.
      (a) S-shaped curve for the fraction of bound pairs, $n(X,N)$ for
      finite DNA chains as indicated. (b) Data collapse for
      finite size scaling for the first order melting as given by the
      Zimm model. Here $\tau=N \ln(X/\mu)\sim \epsilon (T_c-T)/T_c,$
      as in Eq. (\ref{eq:43}).}
    \label{fig:frac}
  \end{figure}
}%

\newcommand{\dpmodel}{%
  \begin{figure}[htbp]
    \centering
    \includegraphics[width=0.4\linewidth,trim= 0pt 0pt 0pt 30pt]{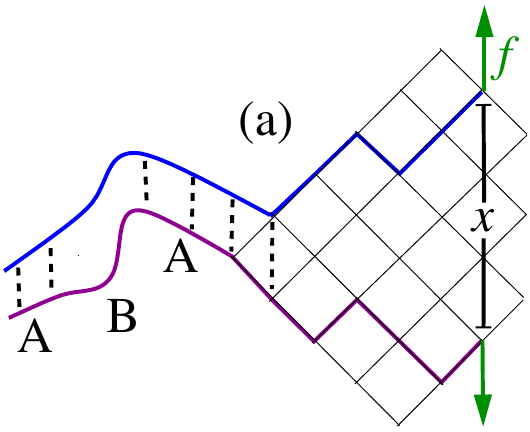}\quad
    \includegraphics[width=0.56\linewidth,trim=20pt 30pt 0pt 0pt]{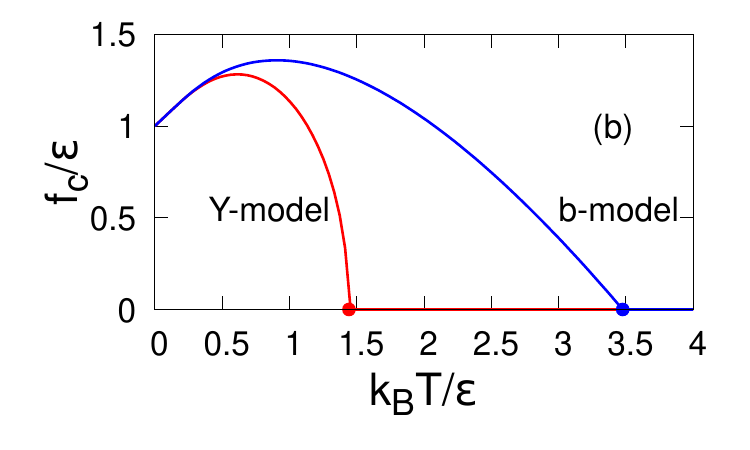}
    \caption{ (a) A two-dimensional DNA model on a square lattice. The
      strands are directed along the diagonal direction (from left to
      right) and are mutually avoiding, meaning they cannot occupy the
      same site or cross each other. At each step, both strands move
      forward by choosing one of two allowed directions, maintaining
      the no-crossing condition.  (b) Phase diagram showing the
      critical unzipping force $f_c(T)$ as a function of temperature
      $k_BT/\epsilon$ for the Y-model and the b-model. The plotted
      curves represent the phase boundaries separating the zipped
      (bound) phase of double-stranded DNA (dsDNA), located below the
      curves, from the unzipped (unbound) phase of single strands,
      located outside. Each curve corresponds to a different model as
      indicated.  }  \label{fig:dpmod}
  \end{figure}
}%

\newcommand{\nvsT}{%
  \begin{figure}[htbp]
    \centering
    \includegraphics[width=1.0\linewidth,trim= 0pt 0pt 0pt
    0pt,clip]{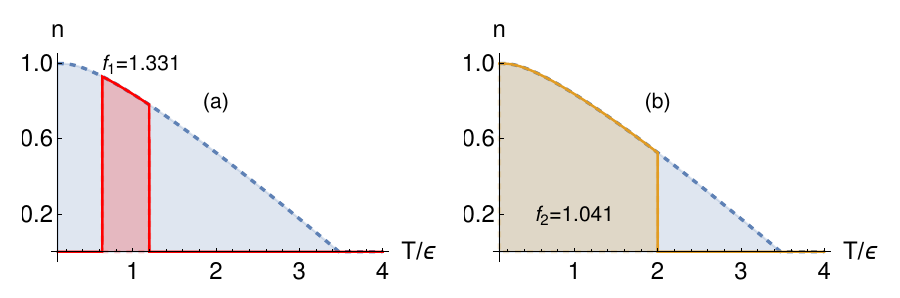}
    \caption{ Melting under force for the b-model. The fraction of
      intact pairs $n(T,f)$ is plotted against temperature $T$ for two
      different forces as indicated. Here $\epsilon=1$. See Fig.
      \ref{fig:dpmod}. The shaded regions under the curves represent
      the bound or zipped phases ($n\neq 0).$ (a) The red curve for
      $f_1=1.331$ shows the fraction as a function of $T$. Note that
      $n=0$ in the left low-$T$ region.  There are two transitions.
      (b) Same as in (a) but for force $f_2=1.041$ which is below the
      reentrant threshold. There is only one transition.  In both
      panels, the blue dashed curve shows the zero-force case as a
      reference. Note that the bound phase profile remains unchanged
      even in the presence of an applied force.  }
    \label{fig:nvst}
  \end{figure}
}%
\section{Introduction}

\subsection{DNA Melting and Unzipping Transitions}
\label{sec:intro}

The importance of DNA in biology became clear in the mid-twentieth century,
with DNA melting, the phenomenon of separating the two strands by
breaking hydrogen bonds, emerging as a key process.  Since the genetic
code is stored in the sequence of bases, strand separation is
essential for both replication, which requires complete dissociation
of strands for genetic information transmission and gene expression,
which involves local strand opening for RNA synthesis \cite{physical}.
Beyond its biological role, DNA melting was also recognized as a novel
type of phase transition, becoming a subject of study in the context
of cooperative phenomena.

\milst

The DNA melting temperature depends on the sequence and environmental
conditions, typically ranging from 40 to 90$^{\circ}$C, or
extreme pH $>9$.  Although strand separation can be achieved {\it in
  vitro} by altering temperature or solvent conditions like pH, such
extreme physical or chemical environments are generally incompatible
with living systems.  {\it In vivo}, this separation is facilitated by
specialized proteins, such as helicases, whose mechanical actions,
like pulling and twisting, actively open up the DNA strands.  These
biological mechanisms can now be replicated and studied using modern
force microscopy techniques.

The idea that an external force can drive a phase transition in DNA,
known as the {\it unzipping transition}, was first proposed in Ref.
\cite{smbjpa}. This force-induced separation of strands is
conceptually distinct from thermally or chemically (e.g., pH) induced
strand separation, which is commonly referred to as {\it melting of
  DNA} \cite{kumarli,vologodskii}. In unzipping, the force is applied
at the open ends (terminals) of the DNA strands, but the transition
itself is not a boundary phenomenon. Rather, it initiates at the end
and propagates along the molecule, particularly evident in the
dynamics of unzipping.  This process stands in contrast to melting,
which involves the disruption of base pairs throughout the DNA.  After
unzipping, the strands are stretched by the applied force, whereas in
melting, they adopt coiled or swollen conformations, typical of
polymers in a good solvent.  Importantly, since polymer stretching
under force does not constitute a phase transition, the thermodynamic
phase of the DNA after unzipping or melting is the same, even though
the mechanisms and conformations involved are different.

The DNA melting and unzipping differ from the standard Landau paradigm
of phase transitions epitomized by the Ising model of 1924
\cite{smb-khare}.  The minimal model studied by Ising consists of two
key terms, (i) an {\it interaction term} that can spontaneously break
the symmetry of the system (i.e., of the Hamiltonian) via temperature
changes, and (ii) an {\it external field term} (e.g., a magnetic
field) that explicitly breaks the same symmetry.  The model was
developed and studied well before the formalization of the concept of
spontaneous symmetry breaking, yet it contained the essential
ingredients that laid the foundation for modern theories of phase
transitions. In contrast, DNA melting, and unzipping involve strand
separation without any obvious underlying symmetry or conventional
order parameter, and the external force driving unzipping does not
couple to symmetry-based order parameter. This paper discusses some
key aspects of these non-Landau transitions.

A study of unzipping timescales \cite{sebastian} soon confirmed the
concept of a force-induced unzipping transition \cite{smbjpa}.  The
transition persists even in heterogeneous sequences \cite{nelson1},
and an experimental phase diagram was later established
\cite{danilowicz,ritort,ansel,heslot}.  Exact solutions of a broad
class of models in various dimensions \cite{smbprl2001,maritan} have
offered further theoretical insights.  Additionally, reentrance
phenomena observed in both exact solutions and numerical simulations
\cite{smbprl2001,smbjpa2001} further reveal the rich and
unconventional nature of the unzipping transition. These features are
explored in some detail in Secs. \ref{sec:unzipping-transition} and
\ref{sec:therm-form-unzipp}.
   
\subsubsection{Outline}
\label{sec:outline}
In Sec. \ref{sec:early-history:-dna}, we present a brief historical
overview of key developments in biology, particularly concerning DNA,
and in contemporary physics, set against the backdrop of the
post-World War II era.  In Sec.  \ref{sec:dna-melting-poland}, we
explore the theory of the DNA melting transition, starting with the
Poland--Scheraga model, which describes melting as a bubble-mediated
process.  Within this framework, both the original Zimm model and the
one-dimensional Ising model \cite{smb-khare} arise as special cases.
The unzipping transition is introduced in Sec.
\ref{sec:unzipping-transition} where it is analyzed through exact
solutions of several models.  Finally, in Sec.
\ref{sec:therm-form-unzipp}, we develop a more general thermodynamic
formulation that extends beyond specific microscopic descriptions,
offering broader insights into the transition. We end with concluding
remarks in Sec. \ref{sec:conclusion}.

\subsection{Early history:  DNA, melting, and molecular biology  }
\label{sec:early-history:-dna}

In the aftermath of World War II, the openness created by the
declassification of wartime scientific research ushered in a new era
of discovery in Physics and Biology.  (see Fig.  \ref{fig:1}). Yet,
this scientific renaissance was marked by a stark contradiction.  The
same state apparatus that facilitated unprecedented innovation also
contributed to large-scale suffering, such as the Bengal famine of
1943.  Engineered by colonial policy and wartime resource
reallocation, the famine resulted in the deaths of an estimated 3
million people \cite{britan}.  This dichotomy invites a sober
reflection on the dual legacies of wartime: scientific advancements
emerging as byproducts of militarized state agendas, and systemic
human tragedies that sustained those very systems.

One immediate outcome of the post-war openness and declassification
was the Lamb shift experiment measuring the
${2}{\textrm{S}}_{\frac{1}{2}}$-${2}{\textrm{P}}_{\frac{1}{2}}$ energy
level splitting in hydrogen.  These experiments were directly enabled
by the advances in microwave technology
\cite{lamb1,lamb11,lamb2,lamb3,lamb4}.  The results marked the formal
beginning of quantum field theory, particularly quantum
electrodynamics (QED) and renormalization.  The success of these
experiments, combined with precision frequency control techniques,
liberated from the constraints of wartime secrecy, laid the essential
groundwork for the invention of the maser and, ultimately, the laser
whose impact on science, technology, and everyday life cannot be
overstated. These breakthroughs stand as examples of what scientific
openness and sustained support can achieve \cite{maser}.

On the Biology front, thanks to the access to isotopes and advanced
instrumentations, the post-war period oversaw landmark discoveries.
These included the identification of DNA as the genetic material
\cite{hershey} and the discovery of its double-helical structure
\cite{watson,franklin}, and paving the way for the emergence of
molecular biology as an interdisciplinary field grounded in the
principles of physics and chemistry.

The Harshey-Chase experiment in 1952 utilized the $\beta$-decay of
$^{35}$S (for proteins) and $^{32}$P (for DNA) to demonstrate that DNA
is the genetic material. This experiment revealed that DNA labeled
with $^{32}$P entered bacteria (specifically, E. coli) when infected
by the bacteriophage T2, while proteins labeled with $^{35}$S remained
outside in the solution \cite{hershey}. The experiment cleverly
employed a kitchen blender, following the use of such blenders (and
their commercial versions) during the war effort to isolate plutonium
\cite{hernandez}.

Nearly a decade earlier, Avery, MacLeod, and McCarty(AMM) had already
used {\it in vitro} methods \cite{avery} to study the propagation of
the virulent strain of bacterial pneumonia, identifying DNA as the
material of inheritance, the so-called ``stuff of life.''  The
findings of AMM were published in the February 1, 1944, issue of the
Journal of Experimental Medicine.  Interestingly, the February 1,
1944, issue of Physical Review featured Onsager's solution to the 2D
square lattice Ising model \cite{onsager,smb-khare}.  While the impact
of Onsager's solution was immediate among physicists, that for the AMM
paper was not so, partly due to the prevailing belief that proteins
were the hereditary material \cite{mccarty,hayes}. It is to be noted
that Onsager and the team of Avery, MacLeod, and McCarty did not
participate in wartime research programs, focusing instead on
fundamental, curiosity-driven science.

With the two nuclear bombings in 1945 came the declaration of the end
of WWII, though with a long tail, with repercussions haunting the
world ever since, particularly in the realm of research.  It is likely
that the attempts to glorify the products of the war efforts cast a
shadow on the AMM work, which failed to get the recognition it
deserved for deacdes \cite{griffith}.

On the physics side, summarising the progress in theoretical physics
during the war, Pauli wrote to Casimir \cite{montroll} that ``nothing
much of interest had happened anyway, apart from Onsager's solution of
the Ising problem.''  However, Guggenheim's paper titled ``The
Principle of Corresponding States''\cite{guggen} of 1945 became a
precursor to the concept of universality, which would emerge as a
dominant theme within two decades of its publication.

Watson and Crick proposed the right-handed double helix model, now
called B-DNA, by combining Franklin's X-ray diffraction data
\cite{franklin} with additional biochemical evidence of equality of
the number of Adenine and Thymine, and Guanine and Cytosine.  There
was no a priori expectation of a preference for either left- or
right-handed forms, because in most inorganic systems and laboratory
processes, mirror symmetry is typically preserved, with both forms of
chiral molecules occurring in equal proportions.  The right-handed
nature of DNA questioned that expectation of symmetry.  The later
discovery of Z-DNA \cite{zdna}, a left-handed form of DNA, as well as
the synthetic production of L-DNA (a mirror image of B-DNA built from
L-enantiomers) demonstrated the possibility of alternative chiral
forms of DNA, even if they are rare or absent in biological systems
\cite{ldna}.  Nevertheless, the B--Z transition (right to left handed)
poses a different type of challenge as the handedness has to change
along the whole length of DNA \cite{bz,bzwang,jmepl}.

Coincidentally, within 3 years of the right-handed double helix
proposal, parity violation, a fundamental asymmetry between a process
and its mirror image, was established in $\beta$-decay (radioactivity)
\cite{leeyang,wuparity}. The inverse process, $\beta^{+}$-decay
(emission of positron and neutrino), also occurs, though rare, and it
violates parity invariance. These findings established that the weak
nuclear force (weak interaction) {\it intrinsically} violates mirror
(parity) symmetry. Despite this insight into fundamental physics and
the eventual success of the Standard Model of particle physics, the
origin of symmetry breaking in biology, particularly in DNA, remains
an enduring enigma.

The double helix paper concluded with an observation, ``It has not
escaped our notice that the specific pairing we have postulated
immediately suggests a possible copying mechanism for the genetic
material'' \cite{watson}.  This statement alluded to the
semi-conservative model of DNA replication, in which each daughter
molecule consists of one parental and one newly synthesized strand.
The model was validated by Meselson and Stahl \cite{meselson}, who
employed non-radioactive heavy nitrogen isotopes ($^{15}$N) to label
the DNA of E. coli. By growing the bacteria in media enriched with
$^{15}$N and subsequently shifting them to a lighter $^{14}$N medium,
they tracked changes in DNA density using equilibrium density-gradient
ultracentrifugation. The resulting intermediate-density bands provided
direct evidence for semiconservative replication, confirming that the
hydrogen bonds between base pairs are broken, and each strand serves
as a template during cell division.

Remarkably, in the same decade, to probe the mechanism of
superconductivity, stable isotopes (such as, $^{202}$Hg, $^{200}$Hg
and others) were used to alter the atomic mass of metals like mercury,
tin, and lead. These mass changes affected the phonon frequencies of
the crystal lattice and, in turn, the superconducting transition
temperature \cite{isotope}. This phenomenon, known as the isotope
effect, provided critical evidence for the role of electron--phonon
interactions in superconductivity and led to the concept of Cooper
pairs, which are bound states of two electrons with opposite spins and
momenta, despite their electrostatic repulsion \cite{cooper}. These
insights culminated in the BCS theory (1957), a milestone in quantum
condensed matter physics \cite{bcs}. This theory clarifies that
superconductivity is an example of {\it spontaneous symmetry breaking}
as opposed to the explicit symmetry breaking in the parity violation
of weak interaction established almost simultaneously.  One may draw
an analogy with magnets, the spontaneous symmetry breaking that gives
us ferromagnets vis-\`a-vis the forced alignment of the magnetic
moments in a paramagnet by an external magnetic field. These are the
two terms of the Ising model mentioned earlier.  Just a year later,
Anderson proposed the localization of all electronic states in
disordered one-dimensional systems, which was a radical idea that
challenged the prevailing belief in the inevitability of extended
electronic states in crystalline solids \cite{anderson}.

The physical separation of DNA strands is only one facet of
replication. The experiments on the semi-conservative mechanism did not
identify the molecular machinery responsible for strand separation or
new strand synthesis for inheritance. The discovery of DNA
polymerases, Pol III in particular, through radioactive nucleotides,
revealed the modularity of DNA synthesis, which turned out to be a
collective process involving multiple enzymes \cite{pol1,pol2}.

The above-mentioned developments in biology and, independently, in
physics reveal a common principle, namely, complex systems often
exhibit emergent behavior that is largely independent of microscopic
details.  Just as DNA replication depends on the general logic of base
pairing rather than the specific chemical identities of the bases,
superconductivity depends on universal features of electron--phonon
interactions rather than the exact details of the lattice structure or
metallic properties.  These insights underscore a unifying theme
across disciplines, that of {\it emergence}; the behavior of complex
systems often arises not from the minutiae of their constituents but
from the structured interactions that give rise to new, collective
laws.  In short, what we discussed are all examples of emergent
phenomena.

What distinguishes double-stranded DNA (dsDNA) from other materials is
its precise Watson--Crick base pairing.  Native base pairs form
hydrogen bonds at strictly complementary positions along the two
antiparallel strands, one running from the $3^{\prime}\to 5^{\prime}$
direction, and the other from $5^{\prime}\to 3^{\prime}$ \cite{sugar}.
This complementarity ensures that the base sequence of one strand
uniquely determines the sequence of the other, introducing a built-in
redundancy crucial for biological processes such as replication, error
correction, and repair. Non-native pairings, in contrast, can lead to
disaster from a biological point of view.

DNA's stability arises from a clear hierarchy of interactions.  Strong
covalent bonds form the sugar-phosphate backbone of each strand, while
significantly weaker hydrogen bonds connect the two strands. Stacking
interactions between adjacent base pairs further stabilize the double
helix, and also produce bending rigidity, measured by a persistence
length. When these native interactions are disrupted, such as during
thermal melting, the double helix separates into two single strands,
each behaving like a flexible polymer. This transition entails a
complete loss of the original rigidity \cite{persist,ha2021,tinland,driessen}.
It is these native interactions that make the melting of DNA a very
special class of problems, even within the broader framework of phase
transitions.

Even before the proposal of the double helix, Thomas, from
UV-absorption experiments in 1951, observed the denaturation of DNA
and predicted the necessity of a secondary structure of DNA
\cite{thomas,thomas1951}.  A confirmation of denaturation came from
Doty and Rice by viscosity measurements \cite{doty}.  The
reversibility of the melting transition was soon established by Murmur
and Doty in 1961, who showed that ``if the thermally denatured DNA
solution is cooled slowly, renaturation of the DNA takes place,
resulting in a restoration of its transforming activity, physical,
chemical and immunological properties.''\cite{murmur}

An Ising-type model to discuss the possibility of a melting phenomenon
was introduced by Zimm in 1960 based on an earlier model by Zimm and
Bragg for protein denaturation \cite{zimm}.  This theory showed that
the melting transition is a sharp first-order transition.

Applying Ising-type models to DNA and protein systems marked a
significant step in developing coarse-grained approaches to biological
problems. A minimal coarse-grained model aims to retain only the
essential degrees of freedom and interactions, such that removing any
component renders the system trivial or unphysical. In the context of
DNA, the primary variable is the formation or disruption of hydrogen
bonds between complementary base pairs; the strong covalent backbone
bonds remain intact during thermal denaturation.  The stacking
interaction, although necessary for stability, is typically treated as
a secondary energetic contribution.

For years, it was believed that the strand separation {\it in vivo}
occurred through passive processes like thermal melting or chemical
instability, but such mechanisms lacked the precision and timing
required for accurate replication. Efforts to find an alternative
mechanism were hindered by the fact that strand separation, unlike
polymerase activity, does not produce easily detectable molecular
products.  The missing link between the semi-conservative mechanism
and DNA synthesis was finally identified in the mid-1970s with the
discovery of the DNA helicase \cite{tuteja,helicase}.

Helicases are molecular machines that couple ATP hydrolysis to
separate the strands in a sequence-independent manner, with their
speed and processivity now well characterized \cite{recbcd,recq,hexaheli}.
Notably, there is not just one helicase. For example, E. coli alone
has around 11 different helicases, each tailored to specific roles. To
ensure helicase function, the unwound strands must be kept apart. This
is accomplished by single-stranded DNA-binding proteins (SSBs),
discovered around the same time.  SSBs bind rapidly to the separated
strands, preventing reannealing and maintaining the single-stranded
template required for polymerase action \cite{mishralevy}. Together,
helicases and SSBs form the core of the replication machinery. Central
to this process is the replication fork, the dynamic junction between
double- and single-stranded DNA. However, helicases and SSBs are just
part of the story, as additional proteins coordinate at the fork to
ensure faithful and efficient DNA replication.

The mechanical action of enzymes such as helicases can be mimicked by
applying external forces to the strands of DNA. This raises a
fundamental question: can double-stranded DNA be separated when held
below its thermal melting point, i.e., unzipped purely by mechanical
force? This phenomenon is known as the unzipping transition. Like
thermal melting, the unzipping transition can be analyzed within a
coarse-grained theoretical framework. Such an approach allows for the
identification of universal or generic features of the transition
independent of microscopic details.

\subsection{Orders of magnitude}
\label{sec:order-magnitudes}

DNA molecules are typically long polymers with $N\sim 10^6$ base pairs
(bp) and, in some organisms, even reaching $10^9$ bp.  Shorter
sequences or oligonucleotides are commonly used in laboratory
experiments.  The large size of DNA makes it well-suited for studying
phase transitions, which are formally defined in the thermodynamic
limit ($N\to\infty$). The base-pairing energetics are central to the
melting behavior of DNA.  Adenine (A) pairs with Thymine (T) via two
hydrogen bonds, while guanine (G) pairs with cytosine (C) via three.
This difference in bonding strength makes GC-rich regions more
thermally stable than AT-rich ones, resulting in higher melting
temperatures.

A simple estimate of the melting temperature can be obtained by
balancing the binding energy and entropic contributions. Taking the
hydrogen bond energy as $\epsilon=10$i kJ/mol and the single-strand
entropy as $s=k_{\textrm{B}} \ln 5$ per base (modeling the strand as a
random walker on a cubic lattice with no immediate reversals,
$k_i{\textrm{B}}$ being the Boltzmann constant), the transition
temperature $T$ is determined by $\epsilon=2Ts$, which yields a
melting temperature of approximately $100^{\circ}$C. While this
overestimates the experimental values, typically 70-90$^{\circ}$C for
AT-rich DNA, the estimate captures the scale properly.

The unzipping force can be estimated from the energy required to break
base pairs (for the bound state) and the work needed to separate the
strands fully (unbound state).  For a bond length of $a\sim
2\times10^{-10}$m, the work done to stretch a chain of length $N$ is
$2 N a f$ which should match the bound state energy $N\epsilon$. The
unzipping force $f$ estimated via $2af=\epsilon$ is $\sim 5-10$ pN.
The force required would be much smaller as one approaches the melting
point because of entropic effects, as we see below.

\section{DNA melting: Bubbles and Junctions}
\label{sec:dna-melting-poland}
From a physical perspective, DNA melting and strand synthesis by DNA
polymerase can be viewed as emergent phenomena that are largely
independent of the specific base sequence. This sequence-independence
allows these processes to operate universally across diverse regions
of the DNA.  A prominent example of this principle is the Polymerase
Chain Reaction (PCR), in which repeated thermal cycling leads to an
exponential amplification (approximately $2^n$ copies after $n$
cycles) of any given target DNA sequence.

The mechanical properties of DNA also reflect this emergent behavior.
The double-stranded form exhibits a bending rigidity not present in
single strands \cite{rigid,palrigidity,debj,debj2}.

In the coarse-grained approach, the double helix is modeled as two
polymers whose configurations contribute significantly to the entropy
of the system.  The monomers in this model are not individual
nucleotides but segments or clusters of bases that behave
collectively. These monomers interact via native base-pairing rules
and are connected by a polymer backbone that remains unbroken
throughout the melting process.

\subsection{The Poland--Scheraga Approach}
\label{sec:poland-scher-model}

The coarse-grained Poland--Scheraga model of DNA melting describes the
molecule as two polymer chains connected by native base pairing
\cite{ps66,fisher,shikha}.  A typical DNA configuration in this
framework consists of alternating bound (helical) and unbound
(denatured loop) segments, separated by junction points, as
schematically illustrated in Fig.~\ref{fig:psbub}.

\psbubble

The partition function of a double-stranded DNA, $Z(N)$, can be
expressed as a sum over all possible configurations consisting of
alternating bound (A-type) and unbound (B-type) segments, with the
junctions contributing a weight $\sigma$.  Therefore,
\begin{equation}
  \label{eq:1}
  Z(N)= \sum_n\sum_{\substack{\{L_j\} \\ \sum_jL_j=N}}
  Z_{\rm A}(L_1)\sigma
  Z_{\textrm{B}}(L_2)\sigma  Z_{\rm A}(L_3)...,
\end{equation}
where $Z_{\alpha}(L)$ is the partition functions of a segment of type
$\alpha$ (A or B) and length $L$.  The outer sum over $n$ accounts for
all possible numbers of segments, and the inner sum runs over all
decompositions of the total length $N$ into alternating segments.  For
$N\to\infty$, the grand partition function becomes
\begin{subequations}
    \begin{align}
 {\cal{Z}}(z)&=\sum_{N=1}^{\infty} Z(N) z^N = C \sum_N
               \left[{\cal{Z}}_A(z)\sigma {\cal{Z}}_B(z) \sigma \right]^N  \label{eq:2}
\\
             &= \frac{C}{1-  {\cal{Z}}_A(z)\sigma^2 {\cal{Z}}_B(z)},\label{eq:32}
     \end{align}
where  ${\cal{Z}}_A(z), {\cal{Z}}_B(z)$ are given by
 \begin{equation}
  \label{eq:3}
   {\cal{Z}}_{\alpha}(z) = \sum_N Z_{\alpha}(N) z^N, \quad (\alpha={\textrm{A, B}}),
 \end{equation}
\end{subequations}
and $C$ is a boundary-dependent factor involving
${\cal{Z}_A},{\cal{Z}_B}$.  The series for
${\cal{Z}},{\cal{Z}_A},{\cal{Z}_B}$ defined above converge within
their respective radii of convergence, determined by the singularities
closest to the origin (dominant singularity) in the complex $z$-plane.
The free energy per base pair is related to the dominant singularity
$z_c$ as
\begin{equation}
  \label{eq:33}
  z_c=\lim_{N\to\infty} [Z(N)]^{-\frac{1}{N}}, {\textrm{ or, }}
  \lim_{N\to\infty}{-\frac{1}{N}} \ln Z(N)=\ln z_c.
\end{equation}
Similarly, the dominant singularities of ${\cal{Z}_A},{\cal{Z}_B}$
determine the respective free energies.

The singularity $z_c$ of ${\cal{Z}}(z)$, Eq. (\ref{eq:3}), may arise
from either ${\cal{Z}}_A(z)$, ${\cal{Z}}_B(z)$, or from
the denominator of Eq. (\ref{eq:2}) when
\begin{equation}
  \label{eq:4}
  1-  {\cal{Z}}_A(z)\sigma^2 {\cal{Z}}_B(z)=0.
\end{equation}
The singularities of ${\cal{Z}_B}, {\cal{Z}_A}$ correspond to the
fully bound and fully unbound phases, respectively, while that of
Eq.~\eqref{eq:4} corresponds to an intermediate, thermodynamically
stable state characterized by finite bubbles.  A phase transition
occurs when the dominant singularity changes due to crossing or
merging of singularities.
 
This is the Poland--Scheraga scheme for studying the melting of DNA and
many other effectively one-dimensional problems \cite{fisher}.

\subsection{ Ising {\it \`a la} Poland--Scheraga: importance of
  junctions}\label{sec:one-dimens-ising}

Let us first apply the above scheme to the Ising case which has a
Hamiltonian $H=-J\sum_i s_i s_{i+1},$ on a one-dimensional lattice
with spins $s_i=\pm 1$.  In this model, the A and B states correspond
to all spins being $+1$ and all spins being $-1$, respectively.  The
parameter $\sigma$ is the Boltzmann factor $\sigma=\exp(-\beta J)$ for
an interface that separates A and B (i.e., $+1$ and $-1$ pair) and
occupies one bond. Since the partition functions for states A and B
are given by $Z_A(N) = Z_B(N) = e^{\beta J N},$ we can express the
grand partition functions as follows:
\begin{equation}
  \label{eq:5}
   {\cal{Z}}_{A,B}(z)= \frac{1}{1-zX}, {\textrm{ where }} X=
   e^{\beta J}.
\end{equation}
The singularity of ${\cal{Z}}(z)$ is then determined by 
\begin{equation}
  \label{eq:6}
   (1-zX)^2= (z\sigma^2),
\end{equation}
where the additional $z$ with $\sigma$ is requiredi,  because the
interface here occupies one bond.   

There are two singularities, $z=(X\pm \sigma)^{-1},$ and the one
closest to the origin is $z_c= (2 \cosh \beta J)^{-1}.$ The free
energy per spin is given by
\begin{equation}
  \label{eq:7}
  \beta f=- \ln 2\cosh\beta J,
\end{equation}
the familiar result obtained by Ising \cite{smb-khare}.  

One may note that the focus on the interface in the Poland--Scheraga
approach to solving the 1-d Ising model is reminiscent of the Landau
argument on the absence of any phase transition in the 1-dimensional
Ising model.

\subsection{The Zimm model for melting: cooperativity}
\label{sec:zimm-model-melting}

An asymmetry in the junction-point weight leads to a first-order
transition in the Zimm model \cite{zimm}, which we discuss next.  In
contrast to the Poland--Scheraga model, the Zimm model for melting
assumes that once the two strands are separated, they cannot rejoin.
This constraint effectively reduces the system to a single-interface
problem, as schematically depicted in Fig.~\ref{fig:psbub}c. Because
of the Y-shaped configurations, the model is also called the Y-model.

In the general case, the rejoining of unbound strands is associated
with a junction weight $\sigma$. When $\sigma=0$ rejoining is
prohibited (left to right in Fig.~\ref{fig:psbub}).  The canonical
partition function for a DNA of length $N$, with a single interface at
position $L$ is given by
\begin{equation}
  \label{eq:8}
  Z(N)= Z_B(N)+ \sum_{L} Z_A(L)\sigma_{+-} Z_B(N-L),
\end{equation}
where $Z_A(L)$ and $Z_B(N - L)$ are the partition functions of the
bound and unbound segments of length $L$ and $N - L$, respectively.
Here, $A$ (or $+1$) represents the \textit{bound} state, and $B$ (or
$-1$) the \textit{unbound} state. The junction weight $\sigma_{+-}$
accounts for the interface between these two domains.
 
The corresponding {grand partition function} is
\begin{equation}
  \label{eq:9}
  {\cal{Z}}(z)={\cal{Z}}_A(z)\sigma_{+-} {\cal{Z}}_B(z),
\end{equation}
where $z$ is the fugacity associated with the chain length.  The
relevant singularity of ${\cal{Z}}(z)$ is the smallest one (in
absolute value) among those of ${\cal{Z}}_{\rm A}(z), {\cal{Z}}_{\rm
  B}(z)$.  A switching of singularities determines the transition from
the bound to the unbound state.

If we take
\begin{equation}
  \label{eq:10}
  Z_A(L)=X^L, {\textrm{ and }} Z_B(L)=\mu^L,
\end{equation}
where the bound state partition function $Z_A$ is purely energetic,
$X=\mu_b e^{\beta \epsilon}$, $-\epsilon (>0)$ being the base pair
energy with $\mu_b$ possible configurations per pair, and the unbound
state partition function $Z_B$ is entropic, $\ln \mu$ being the
entropy per monomer.  The grand partition functions are given by
\begin{subequations}
  \begin{equation}
  \label{eq:11}
{\cal{Z}}_A(z)=(1-zX)^{-1}, {\textrm{ and }} {\cal{Z}}_B(z)=(1-\mu z)^{-1}.  
\end{equation}
The singularities are located at 
\begin{equation}
  \label{eq:15}
  z_a=X^{-1}, {\textrm{ and }} z_b= \mu^{-1}.  
\end{equation}
\end{subequations}
The {phase transition} occurs when these singularities coincide, i.e.,
$X = \mu$, marking the melting point. For $X > \mu$, the bound state
dominates; for $X < \mu$, the system is in the unbound state. This
switch defines a {first-order transition} due to the discontinuity in
the derivative of the free energy.
 
\subsubsection{Transfer matrix}
\label{sec:transfer-matrix}

The connection between the Zimm model and the usual Ising model is via
the transfer matrix.  In fact, the Zimm model is typically solved
using the transfer matrix method.  In analogy with the Ising model,
the transfer matrix, as introduced by Ising, encodes the statistical
weights for transitions between successive states (bound or unbound)
along the DNA strand.  Based on the definition in Eq.~(\ref{eq:8}),
the following weights are assigned:
\begin{enumerate}
\item $X, (1\leq X \le \infty)$ for a bound pair, $++$, 
\item  $\mu (>1)$ for  a $(+-)$ interface ($\sigma_{+-}=1$),  
\item  $\mu$ for an unbound pair $(--)$,
\item $0$ for the $-+$ junction. 
\end{enumerate}
as rejoining is not allowed.   For generality, we may  keep  $\sigma X$
  for the $-+$ junction, taking $\sigma\to0$ at the end.
The transfer matrix is
$$\begin{pmatrix}
  X & \sigma X\\ 
  \mu & \mu
\end{pmatrix}.$$ For $\sigma=0$, the eigenvalues are $X$ and $\mu$.
The largest eigenvalue
\begin{equation}
  \label{eq:12}
  \lambda_{\textrm{max}}= \max(X, \mu),
\end{equation}
determines the state of the system so that a melting transition takes
place from a bound state (for $X>\mu$) to the unbound state (for
$\mu>X$) at
\begin{equation}
  \label{eq:37}
  T_c= \frac{\epsilon}{k_B\ln(\mu/\mu_b)}, \,{\textrm{ where }} X=\mu_b e^{\beta\epsilon}.   
\end{equation}
as defined earlier.  As expected for a one-dimensional system, there
is no phase transition for $\sigma\neq 0$.  A transition occurs in the
limit where the Perron--Forbenius condition gets violated.

\subsubsection{Finite size scaling}
\label{sec:finite-size-scaling}

The Zimm model corresponds to an all-or-none model, also called the
Y-model (because of the Y-like configurations).  This is also called
the zipper model \cite{kittel,nagle} For a finite number $N$, the
partition function is given by $Z=X^N+\mu^N$ and the fraction of bound
pairs, $n$, is given by
\begin{subequations}
\begin{eqnarray}
  \label{eq:13}
  n &=&\frac{X}{N} \frac{\partial\phantom{Z}}{\partial X} \ln Z\\
    &=&\frac{X^N}{X^N+\mu^N}\\
   & {\underset{{N\to\infty}}{=}}&  
     \left\{ \begin{array}{ll}
                   0 & {\textrm{ for }} \mu>X\; ({\textrm{high temperature}}),\\
                   1 & {\textrm{ for }} X>\mu\; ({\textrm{low temperature}}),
              \end{array}\right. 
\end{eqnarray}
with $n=1/2$ at the melting point $X=\mu$. 

In most practical cases involving short DNA chains, the melting curve
displays a characteristic sigmoidal (S-shaped) profile, as illustrated
in Fig.~\ref{fig:frac}a.  Consequently, the point at which $n = 1/2$
is commonly adopted as the \textit{operational definition} of the
melting point, which in this model is a first-order melting
transition.

\sshape

There is, however, a finite-size scaling form expressed in terms of
$\tau=N \ln(X/\mu)$ with $\ln(X/\mu)\approx \epsilon (T_c-T)/T_c$ as
\begin{equation}
  \label{eq:43}
       n(X,N) = \frac{e^{\tau}}{1+e^{\tau}}, ({\textrm{ for large }} N) . 
\end{equation}
\end{subequations}
See Fig. \ref{fig:frac}b. The phase transition occurs only in the
limit $\sigma\to 0$.  For any nonzero $\sigma$, the chain would remain
in the bound state ($n\neq 0$, obtained from the largest eigenvalue of
the transfer matrix) though small bubbles would appear along the
chain.  For this reason, $\sigma$ is referred to as the
`cooperativity' factor \cite{zocchi}.

\subsection{Continuous vs  first-order melting transition}
\label{sec:continuous-melting}

A continuous melting transition is possible if bubbles (B-type) can
form inside the bound state (A-type), in contrast to the all-or-none
situation of the Zimm model. This requires an additional entropic
contribution beyond the extensive term, $\ln Z_B \propto L$, in Eq.
(\ref{eq:10}).  Such contributions arise from correlations that
develop within the bubbles, leading to a power-law dependence on $L$,
viz.
\begin{equation}
  \label{eq:30}
  Z_B(L)=\frac{\mu^L}{L^{\psi}}, {\textrm{ or }} \ln Z_B=L \ln
  \mu-\psi \ln L.
\end{equation}
To motivate such a form, consider modeling each strand as a random
walker. A bubble then corresponds to two such walkers starting at the
origin and reuniting anywhere in space after $L$ steps.  The
probability of reunion for noninteracting walkers in $d$-dimensions
scales as $L^{-d/2}$, i.e., $\psi=d/2$.

In this Poland--Scheraga scheme, a continuous transition is found for
$1<\psi<2$ while a first-order transition for $\psi>2$.  There is no
transition for $\psi<1$ when the chains remain in the bound state at
all temperatures.  For example, for two gaussian chains with native
base pairings, one observes a first-order transition in $d>4$ and a
continuous transition in $2<d<4$.

As an aside, it is worth noting that the configurations of DNA strands
are formally equivalent to those arising in the path integral
formulation of the quantum behaviour of two particles interacting with
a short range potential in $d$ dimensions with the ``time'' variable
corresponding to the position along the strand (i.e., base locations).
The native base pairing corresponds to an equal-time interaction
between the particles, whereas the formation of bubbles corresponds to
excursions outside the potential well, i.e., analogous to quantum
tunneling in the classically forbidden area.  In this gaussian polymer
representation of the quantum mechanical path integrals, one has
$\psi=d/2,$ \cite{smsmb}.  The Poland--Scheraga scheme thus predicts a
bound state for any short-range binding potential in $d=1$, consistent
with the basic results of quantum mechanics. The mapping also extends
to higher dimensions; for $2<d<4,$ a finite melting temperature
corresponds to a critical potential strength required for the
formation of a bound state. Importantly, the ground state energy of
the quantum system maps onto the free energy of the DNA strands.

For $\psi>2$, the average bubble size remains finite, and the melting
transition is first order.  There are arguments \cite{kafri,stella}
that excluded volume interactions in three dimensions can drive the
melting transition first-order. More recent studies based on PERM
simulations indicate that large bubbles are not significant enough to
contribute to the vanishing of rigidity at the melting point
\cite{debj}. In other words, typical configurations for dsDNA consist
of Y-type configurations.  Opening of a Y region is expected to be
easier for dsDNA with an open end, but would be difficult to achieve
for a closed DNA like a plasmid.  Brownian dynamics studies showed
that even though small bubbles do form away from the melting point but
ultimately a large bubble drives the melting transition \cite{soura}.

\subsubsection{Models}
\label{sec:peyr-bish-daux}

Recent neutron scattering experiments have shown that the thermal
behaviour of DNA can be described by only a few parameters reinforcing
the importance of coarse-grained approaches \cite{melt1,melt2,melt3}.
Atomic force microscopy provided details about the conformations,
especially the bubbles \cite{yan,dietler}.%

Solvent conditions (like ethanol-water, ethylene glycol, etc.) and
additives (like formamide, Dimethyl sulfoxide, different salts, etc.)
affect base pairing and stacking by altering electrostatic
interactions, hydrogen bonding, and hydration.  These, in turn,
modulate helix stability.  From a coarse-grained perspective, these
interactions collectively define the effective pairing interaction.
Investigating these factors in various solvents and with different
additives is expected to help establish connections between
coarse-grained parameters and more microscopic interactions
\cite{melt4,melt5,melt6,labchip}.%

Gaussian polymer models with native base-pair interactions, as well as
models incorporating excluded volume effects (self- and mutually
avoiding polymers) have already been discussed.  In the next section,
we will introduce a few lattice-based models.  Another distinct,
effectively one-dimensional model that explicitly incorporates both
hydrogen bonding and the discrete nature of base pair sequences in a
Hamiltonian is the Peyrard--Bishop--Dauxois (PBD) model
\cite{pbd1,pbd2,pbd3,pbd4,ysingh}.  Notably, unlike the gaussian
polymer model discussed after Eq.~\eqref{eq:30}, the PBD model
exhibits a thermal melting transition even in one dimension.

In the PBD model, a variable $y_n \in (-\infty, \infty) $ denotes the
separation between the two bases of the $n$th base pair. The total
energy of a DNA segment is given by the Hamiltonian (kinetic energy is
omitted here),
\begin{subequations}
\begin{align}
  H &= \sum_{n=1}^{N} \left[ V(y_n) + W(y_n, y_{n+1}) \right], \label{eq:51} \\
  V(y) &= V_0 \left(e^{-\alpha y} - 1\right)^2, \label{eq:52} \\
  W(y_1, y_2) &= \frac{1}{2} K \left[1 + \rho e^{-\alpha(y_1 +
      y_2)}\right] (y_1 - y_2)^2. \label{eq:53} 
\end{align}
\end{subequations}
Here, $V(y)$ is the Morse potential mimicking the base-pair binding
energy with $\alpha^{-1}$ determining the width of the potential,
while $W(y_1, y_2)$ models the stacking interaction between successive
base pairs using a modified harmonic potential.  The coupling constant
$K$ governs the stiffness of the interaction, and the parameter $\rho$
enhances the stiffness when the base pairs are in the bound state.  

The Morse potential penalizes negative values of $y$, has a minimum at
$y = 0$ (representing the bound state for $y\approx 0$), and saturates
at $V \to V_0$ for $y \gg 1/\alpha$ (representing the unbound state).
The stacking interaction also weakens as the base pairs separate,
since the exponential term $e^{-\alpha(y_1 + y_2)}$ becomes negligible
for large $y_1, y_2$, effectively reducing the spring constant.

The potential $V(y)$ in Eq. (\ref{eq:52}) effectively restricts motion
to $y\neq 0$ whereas other models discussed earlier allow access to
the full real line.  In quantum terms, while any attractive potential
can bind a particle in one-dimensional free space, confining the
system to a half-line (as in the PBD model) requires a critical
potential strength. Therefore a melting transition is expected by
tuning the parameters like $V_0$ and $\alpha$.

A \textit{kink} in the chain arises from a crossover between regions
of small and large $y_n,$ analogous to the \textit{loop or domain
  boundary} in the Poland--Scheraga model.  In this picture,
denaturation bubbles correspond to excursions of $y_n > 0$, forming a
\textit{constrained random walk} that starts at the origin (intact
base pair), remains nonnegative (physically meaningful separation),
and eventually returns to zero (repaired base pair).  The positivity
constraint distinguishes the model from the Gaussian model.  The
{method of images} can be used to determine the loop entropy exponent
which is given by $\psi = 3/2 $ in one dimension \cite{fisher}. Such a
value of the exponent in the Poland--Scheraga scheme indicates a
\textit{second-order phase transition} at the melting point.

By adjusting the stacking parameter $\rho$, one can sharpen the
transition. However, the transition remains asymptotically
second-order, regardless of the value of $\rho$, though, in a wider
temperature range, the transition may look first-order \cite{pbd2}.

\subsubsection{Continuous  melting}
\label{sec:cont-trans}

In the Poland--Scheraga scheme, the continuous transition takes place
by a gradual increase of the average bubble size.  A continuous
transition is characterized by a diverging length scale
\begin{equation}
  \label{eq:39}
\xi_{\parallel}\sim |t|^{-\zeta}, {\textrm{ as }}
t=\frac{T-T_{\textrm{M}}}{T_{\textrm{M}}}\to 0,  
\end{equation}
in this case, along the chain, such that close to the transition, the
free energy per unit length can be written as
\begin{equation}
  \label{eq:35}
G(t)\sim \frac{k_BT_{\textrm{M}}}{\xi_{\parallel}},
\end{equation}
because $\xi_{\parallel}$ is the relevant length scale at that
temperature.  As polymers, the excursion in space for a segment of
length ${\xi_{\parallel}}$ is
\begin{equation}
  \label{eq:41}
 {\xi_{\perp}}\sim {\xi_{\parallel}}^{\nu}, 
\end{equation}
where $\nu$ is the usual polymer size exponent \cite{flory}.

A consequence of the fact that ${\xi_{\parallel}}$ is the only scale
that matters is a fluctuation-induced attraction when a third strand
is added. In the Poland--Scheraga picture, we take two chains at an
average separation of $R$ in the transverse direction, which can be
taken as the $d$-dimensional space with the chain length as the time
like direction, then a third strand that can form bound pairs with
each of the two strands would produce a change in the free energy as
\begin{equation}
  \label{eq:40}
  \Delta G = - \frac{k_BT_{\textrm{M}}}{\xi_{\parallel}}\, {\cal{G}}(R/\xi_{\perp}),
\end{equation}
because on dimensional ground, $R$ must appear in the above equation
in a dimensionless form.  At the melting point,
$\xi_{\parallel},\xi_{\perp}\to \infty$, and for these scales to drop
out, we need ${\cal{G}}(y)\sim y^p,$ as $y\to 0$ where $p$ is to be
determined with the help of Eq. (\ref{eq:41}).  We then find
$p=-1/\nu$, so that
\begin{equation}
  \label{eq:42}
  \Delta G \sim - \frac{1}{R^{1/\nu}}.
\end{equation}
For gaussian polymers, $\nu=1/2$, and we see an attractive $1/R^2$
interaction in the three-chain system at the critical melting point
which would then lead to a bound state
\cite{jm,tpal,tpal2,mura,damien}.  This is the scaling argument for
the emergence of a $1/R^2$ interaction for the Efimov effect in
three-body quantum mechanics, which predicts a three-body bound state
where no two are bound \cite{braaten}.

\section{Unzipping transition}
\label{sec:unzipping-transition}

The melting temperature is too high to be useful for the opening of
strands at physiological conditions.  This mandates alternative
mechanisms for strand separation or even a part of it as needed in
gene expression (protein synthesis from the DNA sequence).  {\it In
  vivo}, this can be achieved through enzymatic mechanoaction of
various proteins (like helicases).

The basic feature is that a pulling force at the open ends of the two
strands of DNA can break the base pairs cooperatively {\it if the
  force exceeds a temperature-dependent critical value}. This
cooperative dissociation is known as the unzipping transition.

In the path integral framework mentioned in Sec.
\ref{sec:cont-trans}, the force enters as an external source term.  In
the corresponding quantum formulation, this force term behaves like an
imaginary vector potential \cite{smbjpa}.  The central question then
becomes: under what conditions can this external potential destroy the
bound state, which corresponds to the double-stranded DNA (dsDNA)
phase? This forms the basis of the unzipping transition in the quantum
mechanical picture \cite{smbjpa,palPT}.

\dpmodel

\subsection{Exact results: phase boundary}
\label{sec:exact-results:-phase}

The unzipping transition can be studied (Sec.
\ref{sec:zimm-model-melting}) by introducing the end point separation
$x$ as a variable.  Let us consider an exactly solvable model
\cite{smbprl2001}.  Two polymers follow the diagonal direction of a
square lattice, as shown in Fig. \ref{fig:dpmod}a, so that the
coordinate in that z-direction always increases. The monomers of two
strands with the same z-coordinates are the native pairs that form a
base pair if separated by a distance $x=1$ along the other diagonal.
The energy is $-\epsilon (\epsilon>0).$ Here, the fully bound chain at
every base pair also has the option of choosing one of the two
possible ways of growing to the right of the figure, which is also the
option for each single strand if not paired \cite{smbprl2001}.
Therefore, $\mu_1=2$ and $\mu_b=2$ (defined after Eq. (\ref{eq:10}).)

Let us first consider the Y-model.  In terms of the singularities,
given by Eq. (\ref{eq:15}) 
\begin{subequations}
\begin{equation}
  \label{eq:16}
  z_a=\frac{1}{2 X}, {\textrm{ and }} z_b= \frac{1}{4}, (X=e^{\beta\epsilon}).
\end{equation}
The melting point is given by $X=2,$ as in Eq. (\ref{eq:37}).  In the
presence of the pulling force, the force-dependent singularity is
given by
\begin{equation}
  \label{eq:17}
 z_c=[2(1+\cosh \beta f)]^{-1}, 
\end{equation}
which comes from the partition function involving the relative
separation $x$. Any separation $x$, on addition of bonds to the
strands, can change to $x\pm 1$ (Boltzmann factor $=\exp(\pm \beta f)$
or stay the same ($=x$) in two possible ways. Consequently, the
partition function is $2+\exp(\beta f)+\exp(-\beta f)$ which we see in
Eq. (\ref{eq:17}).  The force-related singularity $z_c$ coincides with
that for the open strands, viz., $z_b$ for $f=0$.  As $|z_c|< |z_b|,$
we see a transition from the bound phase controlled by $z_a$ to the
unzipped strands, albeit stretched by the pulling force, only if $f$
is greater than a critical force $f_c(T)$ given by
\begin{equation}
   \label{eq:19}
   f_c(T)= k_BT \cosh^{-1} (X-1),\qquad   {\textrm{(Y-model).}}
\end{equation}
The phase diagram is shown in Fig. \ref{fig:dpmod}b, which shows the
first-order unzipping transition with the coloured region as the bound
or the zipped phase of DNA.  The phase boundary has an infinite slope
at the zero-force melting point $T_c=\frac{1}{k_B\ln 2}$. As pointed
out earlier, the zero-force melting is first-order in this Y-model.

The model that allows bubbles along the length (i.e., not restricted
to the Y-type configurations only), but the strands still mutually
avoiding, undergo a continuous melting transition. This model is
called the b-model. The bound state, unlike the Y-model, now contains
bubbles of denatured regions and is controlled by the singularity
[compare with Eq. (\ref{eq:16})]
\begin{equation}
  \label{eq:36}
  z_a=\sqrt{\frac{X-1}{X}}-{\frac{X-1}{X}}.
\end{equation}
The phase boundary for the b-model is given by
$z_a=z_c$, which gives
\begin{equation}
  \label{eq:18}
    f_c(T)= k_BT \cosh^{-1}\left(\frac{1}{2}
      \frac{1}{\sqrt{1-e^{-\beta\epsilon}}-1+e^{-\beta\epsilon}} -
      1\right ),
 \end{equation}
\end{subequations}
with a zero-force continuous melting at temperature $T_c=
\frac{\epsilon}{k_B \ln(4/3)}$.  See Fig. \ref{fig:dpmod}b for the
phase diagram.

The Y-model is the lattice analog to the Zimm model of Sec.
\ref{sec:zimm-model-melting}, and it exhibits the all-or-none first
order transition.  On the other hand, the b-model that admits
bubble-like excitations along the DNA strands shows a continuous
melting transition. This behaviour is in agreement with the discussion
of Sec. \ref{sec:continuous-melting}, because the reunion exponent
$\psi$, Eq. (\ref{eq:30}), known exactly from renormalization group
(RG) \cite{smsmb}, is $\psi=3/2$, which places the system in the
regime of continuous transitions.

{\it The unzipping transition is first order.} This can be verified by
comparing the fraction of bound pairs, $n$ defined in Eq.
(\ref{eq:13}), on the two sides of the phase boundary. On the unzipped
side, $n=0$ but $n(f,T)\neq 0$ as determined from $z_a$ of Eq.
(\ref{eq:16}) (for the Y-model) or Eq. (\ref{eq:36}) (for the
b-model).

Higher dimensional generalizations of these models have also been
solved exactly, and the first-order unzipping phase boundary for
hypercubic lattices is known \cite{maritan}.  DNA models with full
self-avoidance and mutual avoidance have been studied using different
methods like exact enumeration, Monte Carlo methods on lattices and
for off-lattice models, confirming the first-order unzipping
transition
\cite{kumarli,garimasoftmatter,murthy,jeff,pui,kafri2,lam2005,kierfeld,prabal,cocco2,cocco3}.
Further generalizations like heterogeneity of sequence
\cite{nelson2,sarkar}, randomness in medium (solvent)
\cite{kapri2005},  and randomness in the force distribution have been
analyzed in some detail \cite{kapri2007a,kapri2007b}.  Other studies
include force vs averaged extension relation for two random
heteropolymer chains \cite{nechaev} in the low temperature region, and
the dependence of the unzipping force on salt concentrations studied
using the Peyrard--Bishop--Dauxois (PBD) model \cite{navin} and in the
path integral approach \cite{amnu}. Adsorbed polymers also show
similar unzipping transition \cite{orland,rensburg,iliev,tabbara}.

Methods like exact enumeration \cite{garimasoftmatter} or PERM
\cite{debj,debj2} explore the equilibrium behaviour.  In contrast, the
dynamics-based methods include Monte Carlo methods for coarse-grained
models, and Brownian dynamics or molecular dynamics for more detailed
trajectories.  Recently, a Gaussian network model has been introduced
to study the melting and unzipping of DNA where all covalent
interactions are replaced by gaussian springs with the breakable bonds
by discrete energies. The pathway of pair breaking is then determined
by the least free energy increase.  This method is computationally
efficient and has been used to study different problems, especially
unzipping \cite{granek2,granek,bidisha,sumitra}.

The DNA unzipping problem can be analyzed in two distinct ensembles,
namely, the fixed force and fixed distance ensembles.  In the
fixed-force ensemble, where the applied force $f$ is held constant,
the ensemble-averaged extension is determined by $f$ as $x_f=X(f)$,
whereas in the fixed-distance ensemble, where the extension $x$ is
fixed, the average force required to maintain it is $f_x=F(x).$
Ensemble equivalence requires the consistency conditions, $x= X(F(x))$
and $f=F(X(f)).$ However, differences arise near phase transitions. In
the fixed-force ensemble, one observes a phase boundary, whereas the
fixed-distance ensemble exhibits a coexistence region, a
characteristic of first-order transitions, corresponding to
simultaneous presence of distinct phases. For more discussions, see
Refs. \cite{kapri06,binder,titantah,supurna}.

\subsection{Low temperature reentrance}
\label{sec:low-temp-reentr}

The phase diagrams in Fig. \ref{fig:dpmod}b reveal a striking
reentrant phenomenon at very low temperatures.  When the applied force
exceeds a threshold value (here $f=\epsilon$), the DNA transitions
back to the unzipped phase upon cooling. This behavior appears
counterintuitive, since the bound state is typically the dominant
low-energy configuration. The reentrance arises from subtle entropy
differences between the bound and unbound states, which become
significant even at low temperatures.

The density profiles shown in Fig.  \ref{fig:nvst}a,b illustrate the
melting transition under two different force conditions.  In
Fig.\ref{fig:nvst}a, for a force $f_1>\epsilon,$ the red curve shows
that the bound phase exists only within the shaded region where $n\neq
0,$ indicating two first-order transitions. At lower force, as shown
in Fig.~\ref{fig:nvst}b for $f_2<\epsilon,$ the bound phase spans the
entire low-temperature region up to the melting point (under force).
Importantly, the nature of the bound state in both cases remains
unchanged from the zero-force condition, despite the applied tension.
This fact will be central to the thermodynamic analysis developed in
the following section.

At $T=0$, entropy plays no role, and the transition is determined
solely by energy balance. The critical force $f_c=\epsilon$ marks the
transition point where the bound state becomes unstable, with complete
unzipping occurring for $f>\epsilon$.

At very low temperatures, the excitations mainly involve broken base
pairs near the endpoint where the force is applied.  Let a fraction
$m$ of base pairs be broken near the endpoint, resulting in a strand
separation given by $x=mNa,$ where $a$ represents the unit length.
Because of low $T$, the force fully stretches the unzipped portion,
and a straight chain has no entropy.  However, the bound state loses
its configurational entropy. Let $\sigma_b$ be the residual entropy of
the bound state per base pair coming from the large number of
polymeric configurations it can take.  This entropy is lost on
stretching of the unzipped chains. The free energy change relative to
the fully bound state can be expressed as
\begin{equation}
  \label{eq:46}
  \Delta F(m)=mN(\epsilon +k_B T  \sigma_b- f a),
\end{equation}
where the positive  entropic term  represents the loss of the
bound state entropy.
Henceforth, we set $a=1$  without loss of generality.
A negative  $\Delta F$ favours bond breaking, so that
we obtain a linear  phase boundary 
\begin{equation}
   \label{eq:47}
   f=\epsilon +  {k_B T} \sigma_b,
\end{equation}
for both the models (Y and b), with the unzipped phase stable to the
left of the line.  The positive slope of the phase boundary,
indicating reentrance, originates from the residual entropy $\sigma_b$
of the ground state (e.g., $\sigma_b=\ln 2$ for the model in Fig.
\ref{fig:psbub}).

\nvsT

The reentrant behavior described above represents a universal
phenomenon that persists across different models across dimensions,
arising from the fundamental competition between the energy cost of
bond breaking, the entropic penalty, and the work done to stretch the
strands \cite{smbprl2001}. This phenomenon has been confirmed through
multiple approaches: exactly solvable models have provided
mathematical proofs of this effect \cite{maritan,binder,kapri2009},
while numerical studies, including Monte Carlo simulations and exact
enumeration techniques have established their persistence in
three-dimensional systems with full excluded volume interactions
\cite{smbjpa2001,garimasoftmatter}.  The robustness of this reentrance
effect across these diverse systems underscores its universal nature
in DNA unzipping thermodynamics, rather than being an artifact of
model simplifications. Experimental proposals have been put forward to
observe reentrance by lowering the melting transition temperature
using additives such as formamide, and subsequently monitoring
conformational changes along carefully chosen thermodynamic paths
\cite{skrent}.

\subsection{Force Localization and the Emergence of the DNA Eye Phase}
\label{sec:force-interior-eye}

The mechanical unzipping of DNA is highly sensitive to the point of
force application along the polymer backbone \cite{kapri2004,kapri06}.
With one end of DNA anchored, the location of the force away from the
open end, parameterized by a fraction $s$ (with $0<s\leq 1$) from the
fixed end, plays a crucial role in determining the resulting phase
behavior.  This geometric control stabilizes a distinct ``eye'' phase
featuring an unzipped bubble flanked by double-stranded regions or
terminating at the anchored end.  The phase is identifiable, because it
occupies a finite fraction of the chain, unlike the transient bubbles
of the Poland--Scheraga model.

For $s<1/2$, the force generates an eye-type phase, which terminates
on one side at the anchored end.  At low temperatures, the zipped
phase would then exhibit a first-order transition to the eye phase.
Hpwever, at higher temperatures, the zipped state opens up only if the
force exceeds the threshold value.  Such a phase diagram will have a
triple point where the eye, zipped, and unzipped phases coexist.  The
existence of the eye phase has been shown by the exact enumeration
technique in more realistic models involving self-avoiding polymers as
strands and incorporating the directional nature of the base pairing
bonds\cite{eyekumar,probkumar,navineye}.

We can rationalize the overall behaviour by the energy-entropy balance
argument discussed in Sec. \ref{sec:low-temp-reentr}.  At zero
temperature ($T = 0$), the DNA's response to an applied force depends
on the fractional position $s$ from the anchored end. For $s < 1/2$,
the force creates a central loop or \textit{eye phase}, with energy
and entropy given by
\begin{equation}
  \label{eq:48}
E = (-f + 2\epsilon)sN, {\textrm{ and }} S = -2sN \sigma_b,
\end{equation}
following the arguments for Eq. (\ref{eq:46}). 
The zipped-to-eye transition occurs when $f >f_c(s,T)$ where
\begin{equation}
  \label{eq:14}
  f_c(s,T)= 2(\epsilon + T \sigma_b),  
\end{equation}
which is twice the unzipping
force at $s = 1$. For $s \geq 1/2$, complete unzipping is possible
even at $T = 0$, with 
\begin{equation}
  \label{eq:34}
  E = N\epsilon - f s N,\,{\textrm{ and }} S = -N \sigma_b + 2(1-s)N \sigma_1,   
\end{equation}
where $\sigma_1 = \ln 2$ being the entropy of a single chain per
monomer. The phase boundary is given by
\begin{equation}
  \label{eq:31}
  f_c(s,T) = [\epsilon + T (\sigma_b -2(1-s)\sigma_1)]/s.  
\end{equation}
A left-right symmetry is expected at $s = 1/2$, and surely the phase
boundary shows no reentrant behavior, in agreement with the exact
results \cite{kapri2004}.

\subsection{Y-fork and the junction}
\label{sec:y-fork-juncion}

Two essential features emerge from the above analysis. First, the
first-order unzipping transition allows for phase coexistence between
zipped and unzipped regions, resulting in a characteristic Y-shaped
configuration. This is particularly evident in the fixed-distance
ensemble, where the fraction of the unzipped chain is determined by
the end-point separation. Such Y-forks are biologically realized by
hexameric helicases like DnaB, which maintain a fixed strand
separation while creating a mobile junction point that functions
similarly to an interface. Nonetheless, this junction serves as the
active site for essential biological processes, including replication
and nascent strand synthesis, with helicase activity providing the
primary cooperative mechanism for coordinating various replication
proteins.

The processive action of a helicase can be modeled as the motion of
the interface analogous to front propagation in the Fisher--Kolmogorov
framework \cite{helicase0,helicase1,helicase2,ananya1,ananya2}.  At
the replication fork, the interface corresponds to the probability
profile $p(s),$ which gives the probability that a base pair at
position $s$ along the DNA is open.  Under a fixed-distance ensemble,
phase coexistence between the zipped and unzipped regions produces a
static front described by
\begin{subequations}
\begin{equation}
  \label{eq:49}
  p(s)= \frac{1}{2}\left[1-\tanh\left(\frac{s-s_0}{w}\right)\right],  
\end{equation}
where $w$ is the width of the front or the interface which is located
at $s_0$.  The processive motion of the helicase is then represented
by a moving front, with the profile
\begin{equation}
  \label{eq:50}
  p(s,t)=\frac{1}{2}\left[1-\tanh\left(\frac{s-s_0-vt}{w}\right)\right],  
\end{equation}
\end{subequations}
where $v$ is the propagation velocity, which may serve as a
quantitative measure of helicase processivity.

Second, the eye phase shows direct correspondence with transcription
bubbles in biological systems. Within these bubbles, RNA polymerase
synthesizes proteins, while its processive movement along the template
strand propels the Y-fork forward, maintaining the dynamic equilibrium
between zipped and unzipped regions. This parallel indicates that the
thermodynamic principles governing the model system likely reflect
fundamental mechanisms operating {\it in vivo}.

\section{Thermodynamic formulation of the unzipping transition}
\label{sec:therm-form-unzipp}

We begin with the hypothesis that the bound state dsDNA does not allow
penetration of small external unzipping force \cite{poulomi2}.  Any
weak perturbation is expected to manifest through a linear response,
characterized by an appropriate response function. The central
assumption here is that this response function is strictly zero in the
bound phase of DNA, indicating complete resistance to infinitesimal
unzipping forces.

For the thermodynamics description, we have the temperature, entropy
pair $(T,S)$ and the force, distance pair, $(f,x)$ as two relevant
pairs to describe the unzipping transition.  The relevant
thermodynamic relation in terms of the ``intensive'' quantities $T, f$
would be
\begin{subequations}
\begin{equation}
  \label{eq:20}
  dG=-S dT - x df,
\end{equation}
which gives
\begin{equation}
  \label{eq:21}
  G(T,f)=G(T,0)-\int_0^f x df.
\end{equation}
\end{subequations}
This form  is useful if we know the separation $x$ as a function of
the pulling force $f$ (the relevant isotherms) at a constant $T$. At
the melting point $T=T_c,$ no pulling force is needed for unzipping
and, therefore, $f(T_c)=0.$ 

The continuity of the free energies requires that at $T=T_c$
\begin{equation}
  \label{eq:22}
  G_z(f,T_c)= G_u(f,T_c),
\end{equation}
where the subscript denotes the zipped (z) or unzipped (u) phase.  By
using Eq. (\ref{eq:21}) we then have
\begin{equation}
  \label{eq:23}
   G_z(T,0)-\int_0^{f_c} x_z df=  G_u(T,0)-w_u(f_c), 
\end{equation}
where,
\begin{equation}
  \label{eq:44}
  w_u(f_c)=\int_0^{f_c} x_u df,
\end{equation}
is the work done in the unzipped phase.  Here, $G_u(T,0)$ is the free
energy of two isolated strands in the absence of any force. As we are
talking of two long polymers, the effect of overlap may not be
ignored, but the possibility of native pairings would be rather small.
So effectively $G_u(T,0)$ can be constructed from the free energies of
individual chains.

Our hypothesis requires $x_z=0$ for all $0<f<f_c$, which implies
$G_z(T,f)= G_z(T,0)$ for $T<T_c$.  Therefore, $ G_z(T,0$ from
Eq. (\ref{eq:23}) can be rewritten as
\begin{eqnarray}
  G_z(T,0) &=&  G_u(T,f) +  w_u(f) - w_u(f_c) \nonumber\\
           &=& G_u(T,f) - \int_f^{f_c} x_u df.  \label{eq:24}
\end{eqnarray}
Close to the melting point where the unzipping force is small, the
linear response form $x_u=\chi_T f$ is applicable which yields
\begin{equation}
  \label{eq:25}
  G_z(T,f)=G_u(T,f) +\frac{1}{2}\chi_T (f^2-f_c^2).
\end{equation}
It follows that for $f<f_c$, the bound state is the stable phase as
its free energy is lower than that of the free strands under force.

The entropy difference follows from the $T$-derivative of the free
energy, Eq. (\ref{eq:25}),  
\begin{equation}
  \label{eq:26}
 \Delta S\equiv  S_z(T,f_c)-S_u(T,f_c)=\chi_T f_c(T) \frac{\partial f_c(T)}{\partial T}.
\end{equation}
This entropy relation helps us in determining the shape of the phase
boundary. Let us take the shape of the unzipping boundary close to melting as
\begin{equation}
  \label{eq:27}
  f_c(T)\sim |T-T_c|^{\kappa}, {\textrm{ for }} f_c(T)\to 0, T\to T_c,
\end{equation}
where $T_c$ is the zero force melting temperature.
Combining Eqs. (\ref{eq:26}) and (\ref{eq:27}), and noting that
\begin{equation}
  \label{eq:38}
  f_c\frac{\partial f_c(T)}{\partial T}\sim |T-T_c|^{2\kappa-1},
\end{equation}
 we find the following special cases.

\paragraph{ ${\bm{ \kappa=\frac{1}{2}}}$}
\label{sec:bm-kappa=frac12}

For $\kappa=\frac{1}{2}$, $\Delta S\neq
0$ as $T\to T_c$.  A jump in entropy at the melting point indicates the
presence of latent heat given by $T_c \Delta S$ and is a signature of
a first order transition. Therefore, if the DNA melting transition is
first-order, the unzipping transition line in the force--temperature
phase diagram will exhibit an infinite slope at the melting point.
Specifically, $f_c\sim \sqrt{|T-T_c|}$ at the zero force melting
point.

This square-root dependence is confirmed by the exact solution for the
Y-model, as given by Eq. (\ref{eq:19}) and illustrated in Fig.
\ref{fig:dpmod}b \cite{smbprl2001}.  The infinite slope at the melting
point emerges from the derivative relation $\frac{d}{dx}\cosh^{-1} x =
(x^2-1)^{-1/2}.$

This behaviour persists in higher dimensions as well.  Exact solutions
for the Y-model analogs on hypercubic lattices in $D$ dimensions also
show square-root behaviour near the melting point (first-order in all
dimensions) \cite{maritan}.  Even when bubbles are allowed, as in the
b-model, the melting transition is first-order for $D\geq 5$.  In
these models, the DNA strands extend along one direction and fluctuate
as random walks in the remaining $d=D-1$ transverse directions.  A
strand is then a $d$-dimensional random walker,  with the remaining
direction determining the length.  A strand thus behaves as a
$d$-dimensional random walker, where the longitudinal axis corresponds
to the contour length. It is known that a random walker is recurrent
(i.e., likely to return to the origin) for $d<4$, allowing bubble
formation. For $d>4$, however, the walk is transient, and bubble
formation is suppressed. As a result, behavior similar to the Y-model
is expected. This reasoning is supported by exact solutions, which
show the square-root behavior of the unzipping phase boundary near the
melting point.

\paragraph{${\bm{ \kappa >1/2}}$}
\label{sec:bm-kappa-12}
For $\kappa>1/2$, $ |T-T_c|^{2\kappa-1}\to 0$ as $T\to T_c-.$ The
entropy remains continuous across the melting point at zero force,
indicating a continuous melting transition. In this case, there is no
single unique value of the exponent $\kappa$; rather, there is a
constraint that $\kappa$ must be greater than $1/2$.

The b-model discussed above with the phase boundary given by Eqs.
(\ref{eq:19}) and (\ref{eq:18}), the melting transition is continuous
(second order), and the critical unzipping force scales linearly as
$f_c\sim (T_c -T).$ This means $\kappa=1$ for the phase boundary at
$T\to T_c.$ A similar behaviour has been observed for the b-model in
$d=3$, which shows a second-order melting transition.  Because of log
corrections at $d=2,4$, one needs a more careful analysis with
modified Eq. (\ref{eq:27}).

\paragraph{General comments}
\label{sec:general-results}

There is a strict lower bound that $\kappa$ cannot be less than $1/2,$
as $\Delta S$ has to be finite.  So far no case of a continuous
melting with $\kappa$ in the range $(\frac{1}{2},1)$ is known.

For a self-avoiding polymer, the scaling of extension under a force
crosses over from a linear behaviour $x\sim f$ at small forces to a
nonlinear form $x\sim f^{(1-\nu)/\nu},$ for moderate forces, where
$\nu$, is the polymer size exponent.  This nonlinear response reduces
to the linear form when $\nu=1/2,$ consistent with standard elasticity
theory. The entropy change $\Delta S$ exhibits a scaling behaviour of
the form $\Delta S \sim |T-T_c|^{(\kappa/\nu)-1}$. This aligns with
the linear response result of Eq. (\ref{eq:26}) for $\nu=1/2$.
However close to the melting point, the critical force is small
because the force vanishes at the melting point. As a result, the
linear response regime becomes the more relevant physical description
in this context.

\section{Dynamics and Hysteresis}
\label{sec:hysteresis}

In biological systems, DNA unzipping is central to several processes as
mentioned earlier. To capture its realistic behavior, equilibrium
information must be complemented by dynamic descriptions of unzipping
under both time-dependent and time-independent conditions
\cite{sebastian,smbprl2001,zhang2011,carlon,metzler,carlon2,liretzloff,zdra,cocco}.

An essential question is the time scale to unzip a dsDNA of length
$N$. The approach of Ref. \cite{sebastian}, which maps the problem to
the dynamics of an adsorbed polymer, reveals that in the bound phase
($f<f_c$), there is an energy barrier whose height scales as
$N(f_c-f)$.  However, the pulling process becomes barrierless for $f>f_c$
implying that unzipping occurs spontaneously.  The critical force
$f_c$ thus marks the transition point where the energy barrier
vanishes. In this regime, the characteristic unzipping time scale is
found to scale as $\tau \sim N^2$.

Another feature is the early time dynamics which has a scaling form
\cite{smbprl2001} as
\begin{equation}
  \label{eq:45}
  m(t)\sim N\, q(t/N^z),
\end{equation}
where $m(t)$ is the number of monomers liberated by time $t$, $z$ is a
dynamic exponent and $q(z)$ is a scaling function. The scaling form
seems to be valid over a wide range of problems.  The $N$-independent
limit of Eq. \ref{eq:45} would be $m\sim t^{1/z}$. On the phase
boundary, if we assume the process to be diffusion limited that a
monomer liberated at the fork junction has to equilibrate by diffusion
over the length $m$, then the diffusion time would be $m^2$, so that
the rate equation can be written as $dm/dt\sim m^{-2}$ which suggests
a form $m\sim t^{1/3}, $ as found in simulations \cite{smbprl2001}.

A particularly important scenario involves first-order phase
transitions under periodic external forcing, where the system exhibits
hysteresis due to finite relaxation times. For a periodically varying
force of the form $f = F \cos(\omega t)$ with $F > f_c$ (where $f_c$
is the equilibrium unzipping force at the working temperature), the
$x$-$f$ relationship forms a {\it hysteresis loop} instead of
retracing the equilibrium isotherm
\cite{hatch,hystgarima,janke,kapri2012,kapri2014,kaprikalyan,hystkapri,palhyst,amit,rakesh,kapri2021,exp1,exp2}.
For such a loop, area $A=\oint x\, df$ represents the entropy
production per cycle, and serves as a measure of energy dissipation.
In the low-frequency regime, the area shows characteristic scaling
behaviour\cite{rkphyst,bkchyst}
\begin{equation}
  \label{eq:29}
 A(\omega, F)= (F-f_c)^{\alpha} \omega^{\beta}, 
\end{equation}
where $\alpha$ and $\beta$ are critical exponents that may depend on
system parameters like the chain length.  Moreover, resonances in the
form of peaks in the area versus $\omega$ curve were observed at the
Rouse frequencies of the strands \cite{kapri2014}. See also Ref.
\cite{abhishek,pupo,bergues,gpss}.  Although the scaling of the area
of the hysteresis loop is well-established, maybe with strong length
dependence, it needs to be seen if any underlying nonequilibrium
criticality is at the heart of such power laws.

One may also interpret the ascending part (and separately the
descending part) of a loop as the nonequilibrium work done to go from
one equilibrium state to another. Such nonequilibrium work are
expected to obey the work theorem or fluctuation theorem
\cite{bochkov1977,jarzynski,bochkovusp,worktheorev,poulomi2},
\begin{equation}
  \label{eq:28}
  \left\langle \exp(-\beta W)\right\rangle = \exp(-\beta \Delta F), {\textrm{
      where }} \Delta F= F_{\textrm{f}}-F_0,
\end{equation}
and $W$ is the work done in a nonequilibrium process from a state
marked $0$ of free energy $F_0$ to a state ``f'' with free energy
$F_{\textrm{f}},$ even though at the end of the process, the system
may not be in equilibrium.  The system started in equilibrium in state
$0$ at inverse temperature $\beta$. The angular bracket refers to
averaging done over repetitions of the nonequilibrium processes under
the same dynamics protocol, i.e., the same time dependence of the
parameters \cite{kapri2012,kaprikalyan,hummer}.  An implementation of
the programme will be to generate the free energy curve as a function
of the force and generate the $x$-$f$ equilibrium isotherms
\cite{kapri2012}.

\section{Conclusion}
\label{sec:conclusion}

This paper provided an overview of the force-induced unzipping
transition of a double-stranded DNA. The melting problem has a long
history, dating back to the discovery of the double-helical structure
of DNA. We briefly revisited some key milestones in that historical
development.

A central framework in this context is the polymer-based theory of
Poland and Scheraga, which has played a significant role in modeling
both thermal melting and force-induced unzipping. We discussed the
limit where this theory maps onto a one-dimensional Ising model,
which, despite reaching its centenary in 2025, still thrives as a
fundamental tool in understanding cooperative transitions.
Interestingly, the 1D Ising model does not exhibit any phase
transition in the absence of loop entropy, emphasizing the importance
of configurational entropy in DNA denaturation.  We explained how the
Zimm model of DNA melting appears as another extreme limiting case
within the same Poland--Scheraga formalism, underscoring the unifying
role of this scheme in various theoretical descriptions of DNA
denaturation.

The force-induced unzipping transition, along with its characteristics
across different model systems, has been discussed with an emphasis on
a thermodynamic framework. The unzipping transition naturally gives
rise to structures such as the Y-fork and stable DNA bubbles (called
"eyes"), which closely resemble the biological intermediates involved
in replication and transcription. The dynamics of unzipping,
particularly about hysteresis behaviour, the fluctuation theorems, and
the motion of the Y-fork, offer insights into the nonequilibrium
response of DNA under mechanical stress. Looking ahead, an important
direction will be to couple the dynamics at the Y-fork with those of
other molecular components, such as helicases, polymerases, and
single-strand binding proteins. This integration would help to better
understand the cooperative and processive behaviours that are central
to complex functions in molecular biology.

\begin{acknowledgments} 
The author thanks Saberi Knakon for help with Fig.1.\\[6pt]
\end{acknowledgments}

{\bf Data availibility statement:} Data for Figs. \ref{fig:frac},
\ref{fig:dpmod} and \ref{fig:nvst}  were
generated from Eqs. (\ref{eq:13}), (\ref{eq:43}),  (\ref{eq:36}),
(\ref{eq:18})  using {\small
  MATHEMATICA} and the data would be available on reasonable request.  
This paper has no associated data.

{\bf Conflict of Interest:} No conflict of interest to declare.


\begin{thebibliography}{99}
\bibitem{lamb11} The Lamb--Retherford experiment on the fine structure
  of hydrogen precisely measured the
  ${2}{\textrm{S}}_{\frac{1}{2}}$-${2}{\textrm{P}}_{\frac{1}{2}}$
  splitting in the hydrogen spectrum. This achievement was made
  possible by techniques developed during the wartime work,
  particularly in microwave radiation and the control of small
  magnetic fields\cite{lamb1,lamb2,lamb3}. The observed shift, now
  known as the Lamb shift, challenged, at that time, the predictions
  of Dirac's theory. \cite{lamb2,lamb3,lamb4}.
\bibitem{lamb3} A few statements from Ref. \cite{lamb2} may be
  recalled here.
  \begin{enumerate}
  \item in Sec 4: ``... we inquired whether microwave techniques,
    which advanced so much during the war, would now permit a
    successful and clear-cut determination of the hydrogen fine
    structure... .''
  \item In Sec. 24: ``The magnetic field was produced by a small
    electro-magnet used in this laboratory during the war for testing
    magnetrons.''
  \end{enumerate}
\bibitem{sugar} The carbon atoms in the sugar ring are numbered
  $1^{\prime}, 2^{\prime}, 3^{\prime}, 4^{\prime}, 5^{\prime},$ where
  prime distinguishes  them from other C.
\end{thebibliography}
\end{document}